\documentclass[10pt,aps,prl,showpacs,superscriptaddress,twocolumn,citeautoscript]{revtex4-1}

\usepackage{graphicx}  % needed for figures
\usepackage{dcolumn}   % needed for some tables
\usepackage{bm}        % for math 
\usepackage{amssymb}  
\usepackage{amsmath}
\usepackage{hyperref,color,braket}
\usepackage{xcolor}
\usepackage{ulem}

\newcommand{\beq}{\begin{equation}}
\newcommand{\eeq}{\end{equation}}

\newcommand{\fr}{\bm{r}}
\newcommand{\fR}{\bm{R}}

\newcommand{\fa}{\bm{a}}
\newcommand{\fb}{\bm{b}}

\begin{document}
\title{Dipolar localization of waves in twisted phononic crystal plates}
\author{Marc Mart\'i-Sabat\'e}
\author{Dani Torrent}
\email{dtorrent@uji.es}
\affiliation{GROC, UJI, Institut de Noves Tecnologies de la Imatge (INIT), Universitat Jaume I, 12071, Castell\'o, (Spain)}
\date{\today}
%%%%%%%%%%%%%%%%%%%%%%%%%%%%%%%%%%%%%%%%%%%%%%%%%%%%%%%%%%%%%%%%%%%%%%%%%%%%%%%%%%%%%%%%%%%

%%%%%%%%%%%%%%%%%%%%%%%%%%%%%%%%%%%%%%%%%%%%%%%%%%%%%%%%%%%%%%%%%%%%%%%%%%%%%%%%%%%%%%%%%%%
\begin{abstract}
The localization of waves in two-dimensional clusters of scatterers arranged in relatively twisted lattices is studied by multiple scattering theory. It is found that, for a given frequency, it is always possible to find localized modes for a discrete set of rotation angles, analogous to the so-called ``magic angles'' recently found in two-dimensional materials like graphene. Similarly, for small rotations of the lattices, a large number of resonant frequencies is found, whose position strongly depends on the rotation angle. Moreover, for angles close to those that make the two lattices commensurable a single mode appears that can be easily tuned by the rotation angle. Unlike other twisted materials, where the properties of the bilayers are mainly explained in terms of the dispersion relation of the individual lattices, the special angles in these clusters happen because of the formation of dipolar scatterers due to the relative rotation between the two lattices, enhancing therefore their interaction. While the presented results are valid for any type of wave, the specific case of flexural waves in thin elastic plates is numerically studied, and the different modes found are comprehensively explained in terms of the interaction between pairs of scatterers. The analysis presented here shows that these structures are promising candidates for the inverse design of tunable wave-trapping devices for classical waves.

\end{abstract}
%%%%%%%%%%%%%%%%%%%%%%%%%%%%%%%%%%%%%%%%%%%%%%%%%%%%%%%%%%%%%%%%%%%%%%%%%%%%%%%%%%%%%%%%%%%
\maketitle
%%%%%%%%%%%%%%%%%%%%%%%%%%%%%%%%%%%%%%%%%%%%%%%%%%%%%%%%%%%%%%%%%%%%%%%%%%%%%%%%%%%%%%%%%%%
Twisted bilayers are a special type of quasi-crystals in which two periodic materials are stacked and rotated a relative angle. In this configuration, the periodicity disappears in general and different patterns, also named ``moir\'e patterns'', are formed as a function of the rotation angle. Only for a set of discrete angles the patterns are periodic and the bilayers are in a commensurable phase\cite{zeller2014possible}.
The study of this type of materials has received a renewed interest due to the discovery of the extraordinary properties of bilayers of graphene twisted small angles\cite{morell2010flat,bistritzer2011moire,san2012non,cao2018unconventional,cao2020strange}, giving birth to ``twistronics''\cite{carr2017twistronics}, a new research field devoted to the study of these structures.

Twisted bilayers have also been studied in the realm of photonics\cite{huang2016localization,wang2020localization,hu2020topological} and phononics\cite{jin2020topological}, and it has also been shown that strong localization regimes occur with strong dependence on the rotation angle.

The theoretical study of these complex systems is mainly done within the framework of periodic materials, in which the eigenmodes of the two lattices are coupled by means of properly defined interaction terms\cite{bouchitte2010homogenization,bistritzer2011moire,tarnopolsky2019origin}, although it is clear that the lattice points defining the two crystals are closer each other in a slightly twisted configuration, for which the interaction between points will be much higher than in the individual lattices. Therefore, a deeper theoretical analysis has to be done, taking into account these stronger interactions.

In this work, we present a comprehensive study of the localization of waves in twisted bilayer crystals. We have shown that multiple scattering between closer scatterers plays a central role in the formation of localized states. It will be shown that, for a given frequency, a set of localized modes appear for a discrete set of rotation angles. Also, for a small rotation angle, it will be shown that a large number of localized modes appear, nonexistent in the periodic or commensurable phases. Numerical examples will be given for the special case of flexural waves in thin elastic plates with spring-mass attachments, since Green's function in this case is non-singular at the origin and allows for a deeper understanding of the underlying physics, although the conclusions can be applied to any kind of waves.

Let us assume we have a cluster of $N$ point scatterers located in positions $\fR_\alpha$, for $\alpha=1,2,\ldots,N$. Each scatterer is defined by a characteristic impedance $t_\alpha$, which can be a resonant quantity, depending on the response model employed. When an external incident field $\psi_0(\fr)$ impinges the cluster, the total field will be the sum of the incident field plus the scattered field by all the scatterers of the cluster\cite{foldy1945multiple,martin2006multiple}, 
\beq
\psi(\fr)=\psi_0(\fr)+\sum_{\alpha=1}^N B_\alpha G(r-\fR_\alpha),
\eeq
where $G(\fr)$ is Green's function of the corresponding wave equation and the coefficients $B_\alpha$ are obtained from the system of equations
\beq
\label{eq:mst}
\sum_{\beta=1}^N M_{\alpha\beta}B_\beta=\psi_0(\fR_\alpha),\quad \alpha=1,2,\ldots,N
\eeq
with the matrix elements $M_{\alpha\beta}$ given by
\beq
\label{eq:M}
M_{\alpha\beta}=\delta_{\alpha\beta}t_\alpha^{-1}-G(\fR_\alpha-\fR_\beta).
\eeq
The eigenmodes of the cluster can be obtained by the imposition of a non-trivial solution of the system of equations \eqref{eq:mst} when there is no incident field $\psi_0$, which is equivalent to impose the determinant of the M matrix to be equal to zero\cite{ochiai2002localized} or, equivalently, that this matrix presents a zero eigenvalue, which is a more suitable condition from the numerical point of view. 
%%%%%%%%%%%%%%%%%%%%%%%%%%%%%%%%%%%%%%%%%%%%%%%%%%%%%%%%%%%%%%%%%%%%%%%%%%%%%%%%%%%%%%%%%%%
\begin{figure}[h!]
	\centering
	\includegraphics[width=\linewidth]{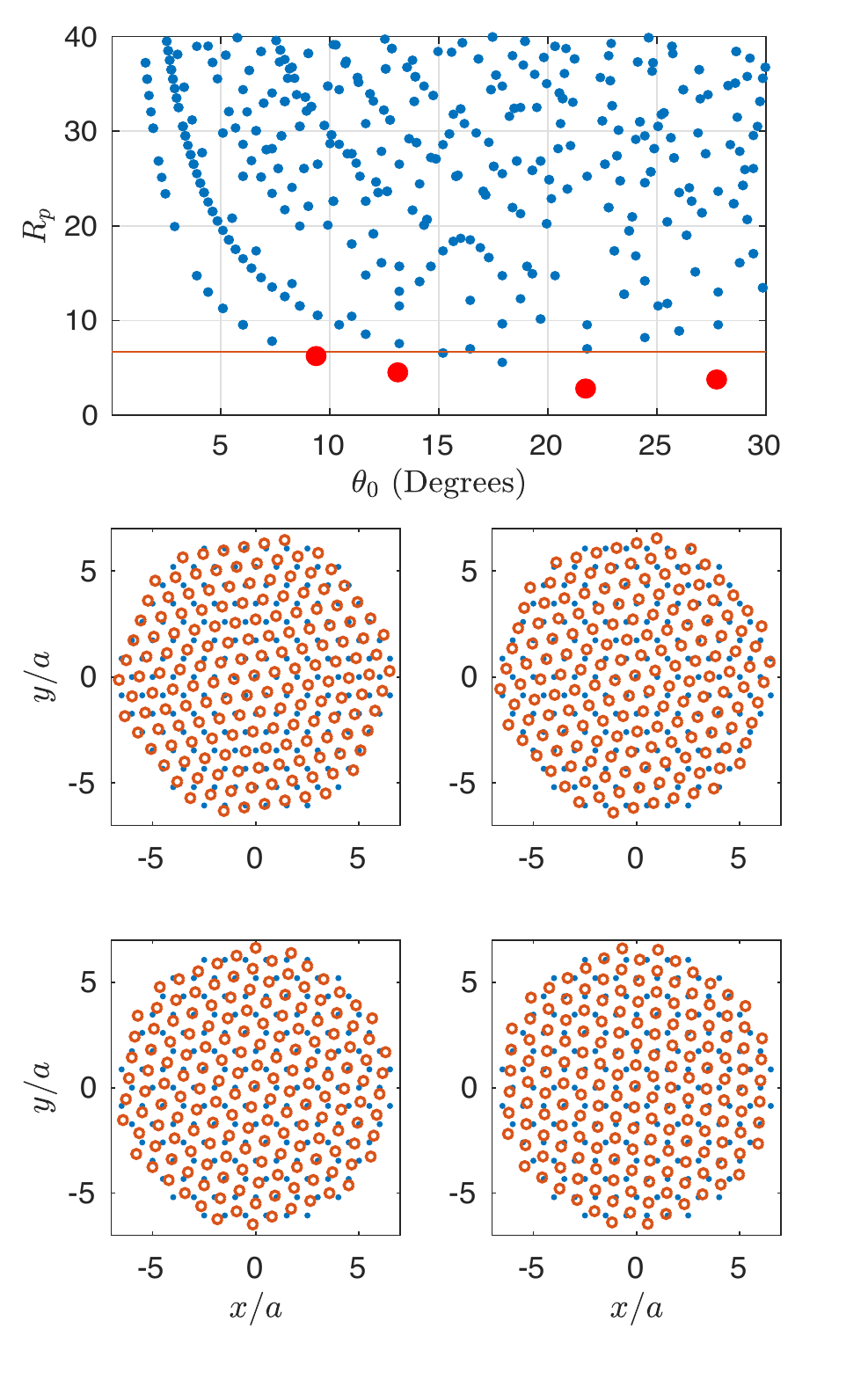}
	\caption{Upper panel: Distance $R_p$ between two equivalent points at a super-lattice as a function of the rotation angle. Red points show the configurations depicted in the clusters of the lower panel, corresponding to the supper-lattices from which the $R_p$ are smaller.}
	\label{fig:sch}
\end{figure}
%%%%%%%%%%%%%%%%%%%%%%%%%%%%%%%%%%%%%%%%%%%%%%%%%%%%%%%%%%%%%%%%%%%%%%%%%%

Let us assume now that the cluster of scatterers is divided in two sets, $a$ and $b$, defined by positions $\fR_a=n_1\fa_1+n_2\fa_2$ and $\fR_b=m_1\fb_1+m_2\fb_2$, with the set $\{\fb_1,\fb_2\}$ being a rotated version of the set $\{\fa_1,\fa_2\}$. For an arbitrary rotation angle $\theta_0$, the two lattices are incommensurable, i.e., we have that $\fR_a\neq\fR_b$ for all the possible lattice points, thus periodicity is broken. But we can find some special values of $\theta_0$ so that, for a given set of $n_1,n_2,m_1,m_2$ we arrive to $\fR_a=\fR_b\equiv\fR_p$, so that periodicity is recovered, being the smaller of $\fR_p$ the new period of the lattice. 

Let us define the complex vector $\fa$ such that $a_\ell=a_{\ell x}+ia_{\ell y}$, for $\ell=1,2$, and similarly for $\fb$. In this representation, we have that $\fb=\fa e^{i\theta_0}$, so that $\fR_b=(m_1\fa_1+m_2\fa_2) e^{i\theta_0}$. The clusters will be in a commensurable phase as long as the rotation angle $\theta_0$ given by 
\beq
\theta_0=i\ln \left[\frac{n_1a_{1x}+n_2a_{2x}+i(n_1a_{1y}+n_2a_{2y})}{m_1a_{1x}+m_2a_{2x}+i(m_1a_{1y}+m_2a_{2y})}\right]
\eeq
is a real quantity.

Figure \ref{fig:sch}, upper panel, shows the period $R_p$  of the super-cell as a function of the rotation angle $\theta_0$ for a triangular arrangement of lattice constant $a=1$. For small angles the period is divergent, then we will need a large cluster of scatterers if we want to have some degree of periodicity. For $R_p<10a$ the number of commensurable angles is small, then building finite clusters below that radius will only have a few commensurable phases. Lower panel of figure \ref{fig:sch} shows a cluster of radius $R_c=6.7a$ at the four commensurable phases marked with red dots in the upper panel. We see how the Moir\'e patterns present a small periodicity. At the lattice points, where the condition $\fR_a=\fR_b$ is met, two scatterers of the different lattices are at the same point.  

The relative rotation of the two clusters allows the formation of a set of ``dipoles'', whose distances will be tuned by means of the rotation angle $\theta_0$. These scatterers will be closer each other than they are in the individual lattices, so that their interaction in this configuration is higher. Since the rotation occurs from the center of the cluster, the relative distance between dipoles depends on the relative position of the scatterers in the cluster, so that in general a large number of different dipoles will be formed and, consequently, a large number of different resonant modes.

In the following lines we will show that these dipoles are the responsible for the strong localization characteristics of twisted lattices. Therefore, it is important to define the ``cluster potential'' $\mathcal{U}$,
\beq
\label{eq:U}
\mathcal{U}=\frac{1}{N_c^2}\sum_{\forall a,b}\frac{1}{|\fR_a-\fR_b|},
\eeq
where $N_c$ is the number of scatterers in one cluster, so that $N=2N_c$. The cluster potential quantifies how close to each other are the scatterers of cluster $a$ and $b$. This quantity has poles at the commensurable angles, i.e., around those angles where a periodic pattern is formed again. 
%%%%%%%%%%%%%%%%%%%%%%%%%%%%%%%%%%%%%%%%%%%%%%%%%%%%%%%%%%%%%%%%%%%%%%%%%%%%%%%%%%%%%%%%%%%

The role of the dipoles in the localization of modes can be understood from the form of the multiple scattering matrix $M$ given by equation \eqref{eq:M}. As we mentioned before, the modes of the cluster are defined by the zeros of the determinant of matrix $M$. In the first approximation we ignore the off-diagonal terms and consider only the diagonal of the matrix, which is the inverse of the impedance term $t_\alpha$, we have then
\beq
|M|\approx \prod_{\alpha=1}^N \frac{1}{t_\alpha}.
\eeq
In this approximation, the modes of the cluster will be defined by the zeros of $t_\alpha^{-1}$, i.e., the poles of $t_\alpha$. This condition might be satisfied for instance if the scatterers are local resonators at their resonant condition.

Let us assume now that we are in a frequency region where no resonance occurs. Resonant modes can happen now due to the interaction between closer scatterers. Given the geometry of the rotated cluster, for small rotation angles a set of $N_c$ dipoles will be formed. We can consider then that matrix $M$ factorizes in a set of $2x2$ matrices $M_d$, so that the modes of the cluster are given by the different modes of each dipole. Thus, dipolar modes are defined by the zeros of the determinant
\beq
|M_d|=(t_0^{-1}-G(\bm{0}))^2-G(\fR_{ab})^2
\eeq
where $\fR_{ab}$ is the distance between the two scatterers forming the dipole, which have been considered identical and with $t_\alpha=t_0$. Since this distance depends on the position of the scatterers in the cluster, for a given rotation angle we will have a set of different resonances.

Numerical experiments can now be done to compare the existence of resonances due to the dipolar interaction. However, the Green function $G(\fr)$ is singular for the Helmholtz equation in 2D, and some normalization methods have to be done to properly\cite{martin2006multiple} analyze these effects. Nevertheless, flexural waves in thin plates of mass density $\rho$, rigidity $D$ and thickness $h$, assuming harmonic time dependence of frequency $\omega$, are described by means of the bi-Heltmholtz equation\cite{norris1995scattering,torrent2014effective}, with a Green's function given by\cite{torrent2013elastic,packo2019inverse}
\beq
G(\fr)=\frac{i}{8k^2}\left(H_0(kr)-H_0(ikr)\right),
\eeq
where $k^4=\omega^2\rho h /D$, which has the remarkable property of being finite at the origin, since $G(\bm{0})=i/8k^2$, being therefore more suitable for numerical calculations. 

Using this model, the point scatterers can be considered spring-mass attachments to a thin elastic plate, with the impedance $t_0$ defined as
\beq
t_0=\gamma_0\frac{\Omega^2\Omega_R^2}{\Omega_R^2-\Omega^2},
\eeq
 with $\Omega=\omega\sqrt{\rho h a^2/D}$ being the reduced frequency and with $\gamma_0$ and $\Omega_R$ being the reduced mass and resonant frequency of the resonator, respectively\cite{xiao2012flexural,torrent2013elastic}. A more accurate description should require the inclusion of some damping in the denominator but we are far away from the resonant condition.

\begin{figure}[ht]
	\centering
	\includegraphics[width=\linewidth]{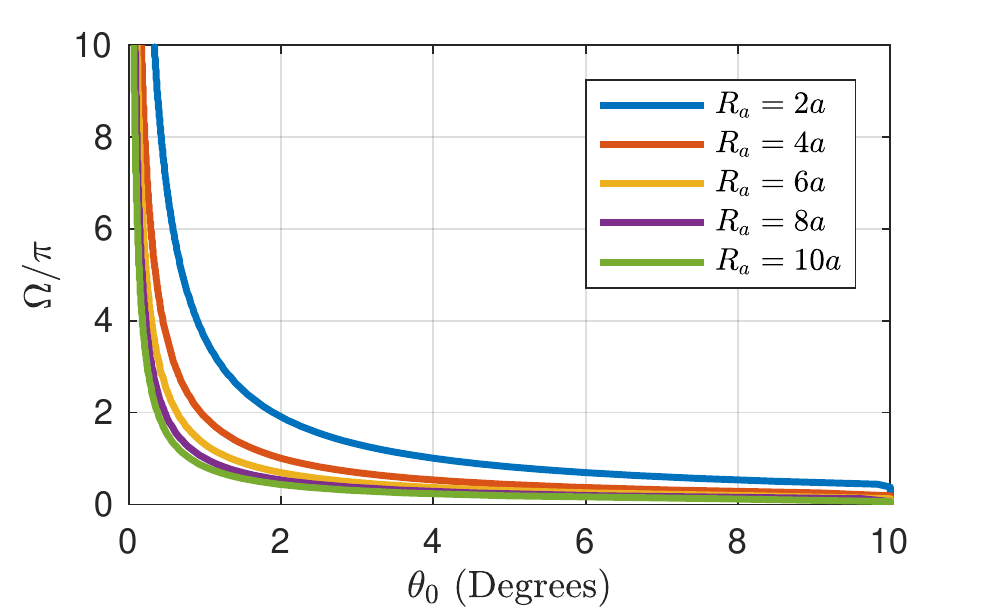}
	\caption{Resonant frequency as a function of the rotation angle for a dipole of scatterers formed at different distances from the center of the cluster.}
	\label{fig:lmin}
\end{figure}

The condition $|M_d|=0$ can only be achieved for complex frequencies, since it is an open system with no bounded modes of infinite lifetime. However, we can see an approximated position for the modes if we plot the smallest eigenvalue of $M_d$ for real frequencies. Then, the normalized frequency $\Omega$ at which the condition $\min{\lambda_d}$ is satisfied is shown in figure \ref{fig:lmin} as a function of the rotation angle $\theta_0$ for different positions of the dipoles in the cluster. We have selected $\gamma_0=200$ and $\Omega_R=20\pi$ in this example. When the rotation angle is large enough the dipoles are too far away to excite a resonance. Even though, the periodic nature of the lattice makes that near the commensurable angles other dipoles will be formed, as shown in figure \ref{fig:sch}, and new modes will appear. 
%%%%%%%%%%%%%%%%%%%%%%%%%%%%%%%%%%%%%%%%%%%%%%%%%%%%%%%%%%%%%%%%%%%%%%%%%%%%%%%%%%%%%%%%%%%
%%%%%%%%%%%%%%%%%%%%%%%%%%%%%%%%%%%%%%%%%%%%%%%%%%%%%%%%%%%%%%%%%%%%%%%%%%

\begin{figure}[h!]
	\centering
	\includegraphics[width=\linewidth]{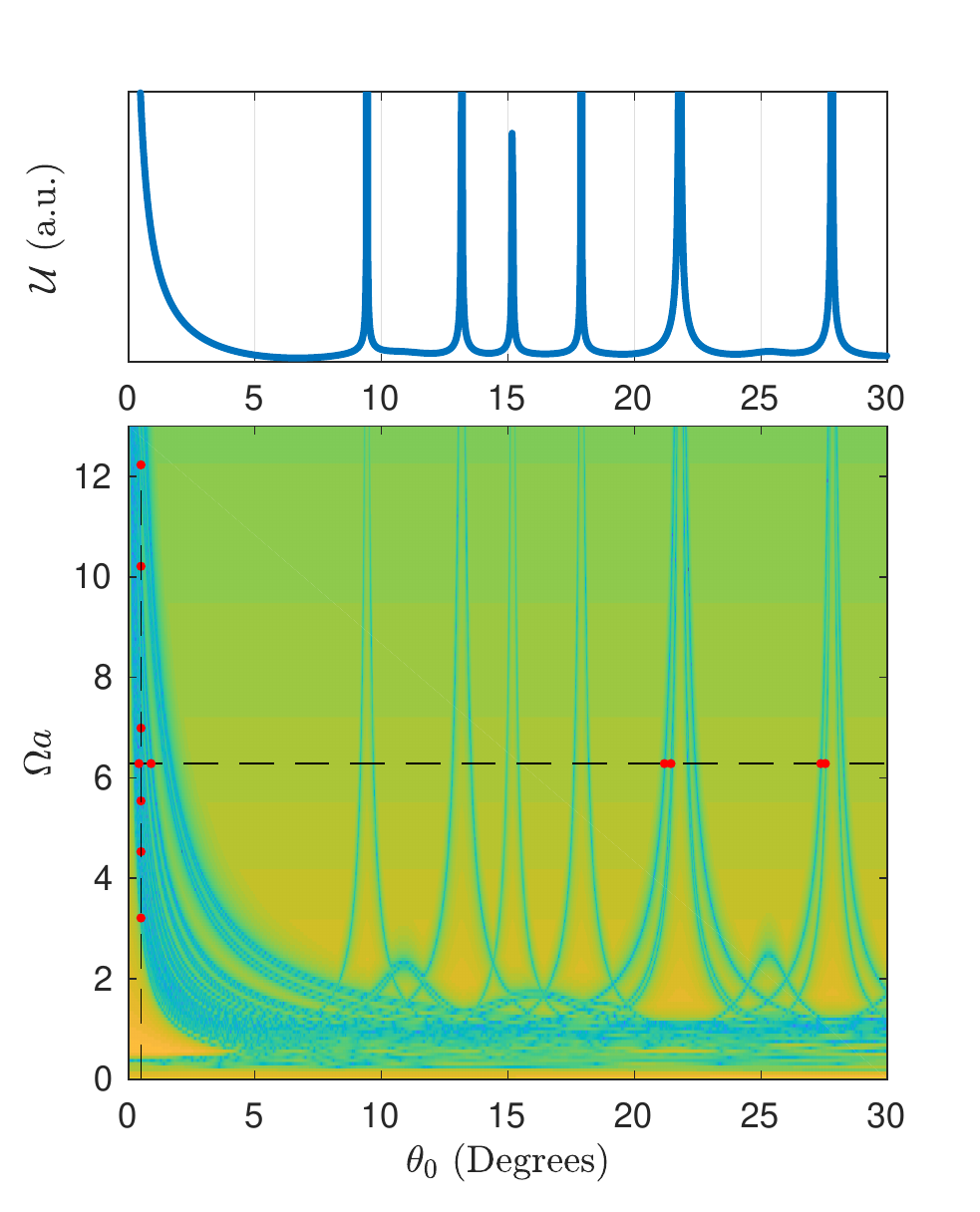}
	\caption{Upper panel: Cluster potential $\mathcal{U}$ as a function of the rotation angle for a circular cluster of radius $R_c=6.7a$ with a triangular lattice. Lower panel: Absolute value of the smallest eigenvalue of the matrix $M$ for the cluster. Blue regions show the smallest value and correspond to the resonances of the cluster. Red dots are the modes depicted in figures \ref{fig:fields_omega} and \ref{fig:fields_theta}}
	\label{fig:triangular}
\end{figure}

\begin{figure}[ht]
	\centering
	\includegraphics[width=\linewidth]{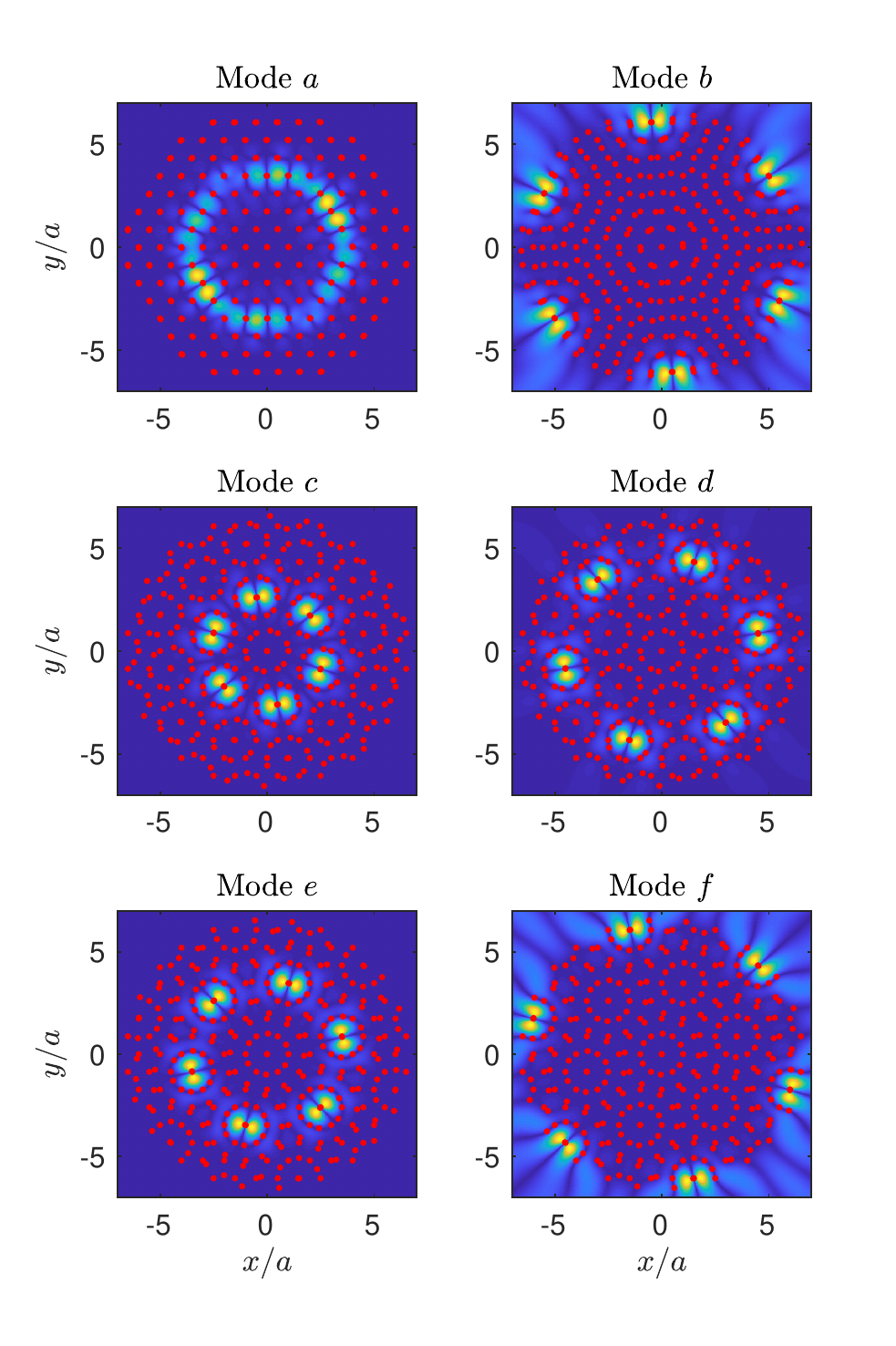}
	\caption{Field patterns corresponding to the red dots of the horizontal line $\Omega a=2\pi$ of figure \ref{fig:triangular}.}
	\label{fig:fields_omega}
\end{figure}

The global behaviour of the cluster can be shown in figure \ref{fig:triangular}, where the smallest eigenvalue of the $M$ matrix of a circular cluster of radius $R_c=6.7a$ and scatterers with the same properties as before has been represented. The underlying lattice of the cluster is triangular and the total number of scatterers in each cluster is $N_c=163$, so the full structure has 326 scatterers. The upper panel shows the cluster potential $\mathcal{U}$ defined in equation \eqref{eq:U}. We can distinguish in this map two types of modes, all of them related with dipolar resonances. The first type of modes is due to the dipoles formed when the cluster is rotated a small angle, so that the scatterers of the two clusters are very close each other. These modes appear in the range $\theta_0=0^\circ$ to $\theta_0=5^\circ$. Above this angle the scatterers of each cluster begin to be too far away each other and dipolar modes are not possible. The second type of modes appears near the peaks of $\mathcal{U}$, where the two lattices form a commensurable structure and, consequently, some of the scatterers at each cluster are at the same position, but not all of them. The modes corresponding to the last two commensurable angles are splitted in two; the reason for this splitting is that at these angles the period of the super-lattice is so mall that there are two periods in the cluster, for a given rotation angle we excite the dipoles of one period, but due to the radial roation, the dipoles of the second period are not excited, since their distance is smaller. As we rotate the angles, these dipoles are finally excited as well. Near the angles $\theta_0=10^\circ,15^\circ$ and $25^\circ$ some modes appear which do not correspond to the peaks of $\mathcal{U}$, but there the dipoles are close enough to excite a resonance.
%%%%%%%%%%%%%%%%%%%%%%%%%%%%%%%%%%%%%%%%%%%%%%%%%%%%%%%%%%%%%%%%%%%%%%%%%%%%%%%%%%%%%%%%%%%
%%%%%%%%%%%%%%%%%%%%%%%%%%%%%%%%%%%%%%%%%%%%%%%%%%%%%%%%%%%%%%%%%%%%%%%%%%
Figure \ref{fig:fields_omega} shows examples of the different modes found along the line $\Omega a=2\pi$, and marked in figure \ref{fig:triangular} as red dots in the horizontal line. Labeled from $a$ to $f$, the patterns correspond to the rotation angles $\theta_0=0.42,9.18,21.18,21.44,27.36,27.54$ degrees. Mode $a$ corresponds to the excitation of dipoles due to a small rotation angle, while all the other modes correspond to the excitation of dipolar modes because of the commensurable phase of the cluster. Modes $c-d$ and $e-f$ are clearly paired and correspond to the excitation of dipoles in the same commensurable phase but at different distances due to the radial symmetry of the rotation, as explained before.

%%%%%%%%%%%%%%%%%%%%%%%%%%%%%%%%%%%%%%%%%%%%%%%%%%%%%%%%%%%%%%%%%%%%%%%%%%%%%%%%%%%%%%%%%%%
%%%%%%%%%%%%%%%%%%%%%%%%%%%%%%%%%%%%%%%%%%%%%%%%%%%%%%%%%%%%%%%%%%%%%%%%%%

\begin{figure}[ht]
	\centering
	\includegraphics[width=\linewidth]{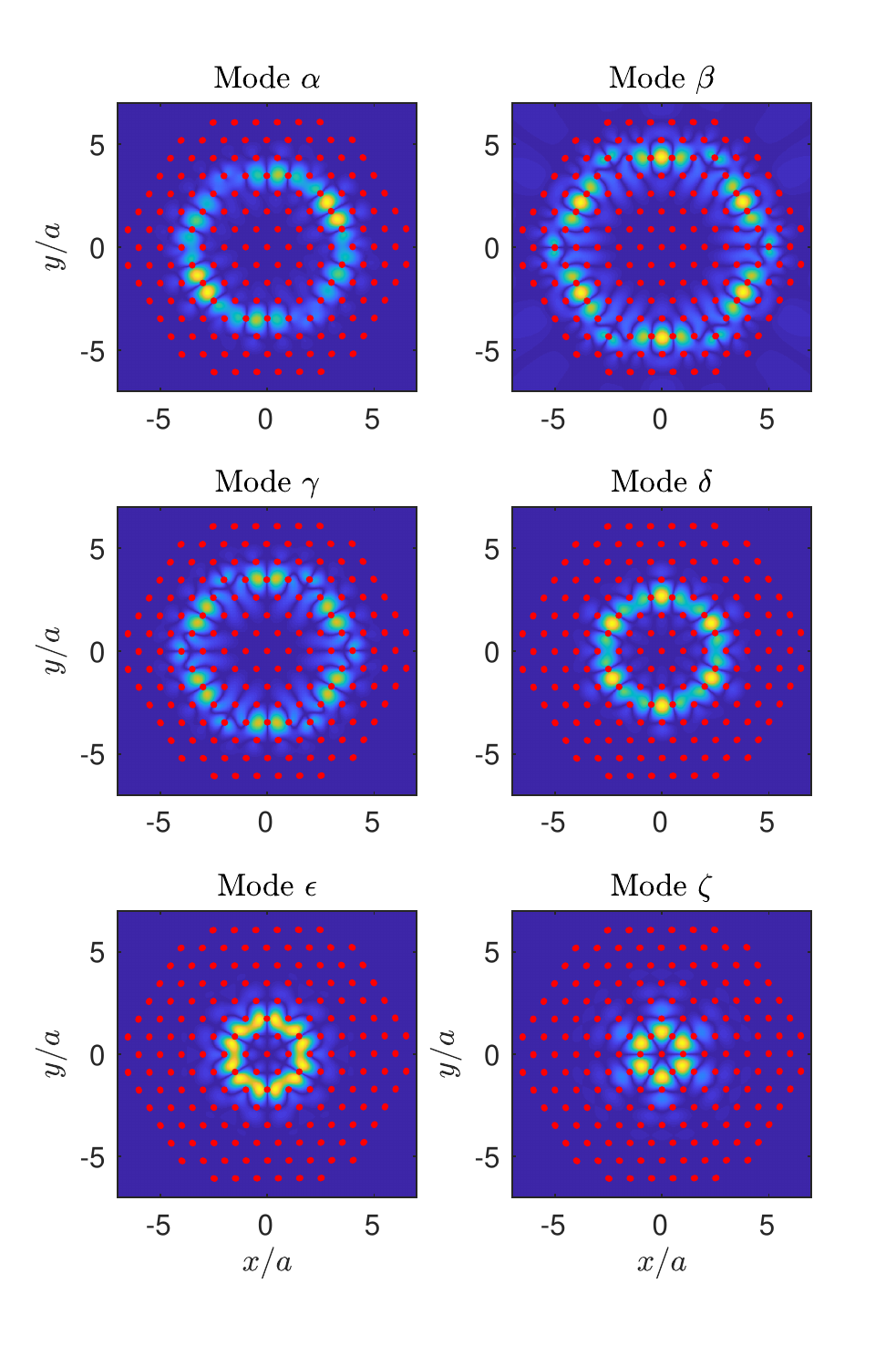}
	\caption{Field patterns corresponding to the red dots of the vertical line $\theta_0 =0.5^\circ$ of figure \ref{fig:triangular}.}
	\label{fig:fields_theta}
\end{figure}

Figure \ref{fig:fields_theta} shows the different modes found at a fixed small rotation angle of $\theta_0=0.5^\circ$, so that a large number of resonances is excited. Labeled from $\alpha$ to $\zeta$, we show plots at frequencies $\Omega a=3.21,4.53,5.54,6.99,10.21,12.23$. The modes are more and more interior to the cluster (except mode $\alpha$, probably a cluster mode). This variation is due to the fact that for a small rotation angle the distance between dipoles is radially increasing, therefore following the curve of figure \ref{fig:lmin} the wavelength of the corresponding mode will also be increasing with the distance to the center or the cluster and the frequency will be decreasing. This remarkable property shows how we can generate a large number of resonant modes for a small rotation angle, in opposite to what we have near the commensurable angles, where the number of frequencies is one or two for small clusters, though tunable with the rotation angle.

%%%%%%%%%%%%%%%%%%%%%%%%
\begin{figure}[h!]
	\centering
	\includegraphics[width=\linewidth]{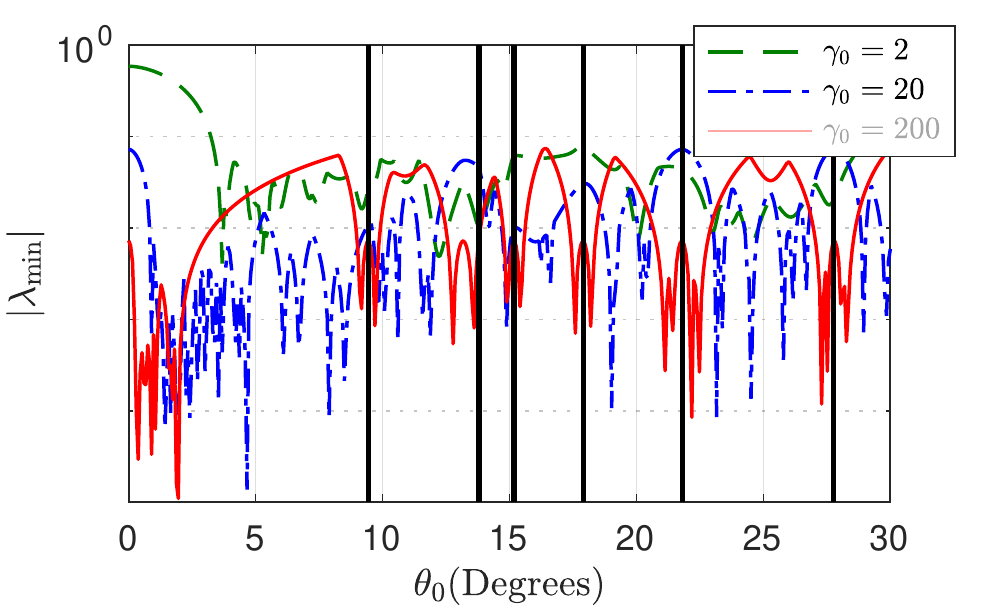}
	\caption{Influence of the impedance parameter in the formation of localized modes.}
	\label{fig:lambda}
\end{figure}

%%%%%%%%%%%%%%%%%%%%%%%%%%%%%%%%%%%%%%%%%%%%%%%%%%%%%%%%%%%%%%%%%%%%%%%%%%
Finally, figure \ref{fig:lambda} shows the effect of the impedance term $\gamma_0$ with the localization and number of modes. Results are shown for $\Omega a=2\pi$. For a high value of $\gamma_0$, which is the situation considered so far, the modes localize at small rotation angles and near the commensurable ones, as has been discussed before. However, decreasing the value of $\gamma_0$ we find that, since the interaction between dipoles is weaker, the modes are more spread along the different angles. While we still can see a large number of these at special values of the rotation angle, their interpretation is however more complex since localization can be due to the interference between scattered waves, in a similar mechanism to ``Anderson localization''\cite{wiersma1997localization}.
%%%%%%%%%%%%%%%%%%%%%%%%%%%%%%%%%%%%%%%%%%%%%%%%%%%%%%%%%%%%%%%%%%%%%%%%%%%%%%%%%%%%%%%%%%%
%%%%%%%%%%%%%%%%%%%%%%%%%%%%%%%%%%%%%%%%%%%%%%%%%%%%%%%%%%%%%%%%%%%%%%%%%%

In summary, it has been found that the multiple scattering of waves between two nearby scatterers creates localized waves, so that when two lattices of scatterers are stacked and relatively twisted a large number of dipoles is formed and, consequently, a set of resonances appears. We have found that, for a fixed frequency, this set of resonances appears at special values of the rotation angle, which have been comprehensively explained in terms of multiple scattering theory . This work shows that, since the scatterers in the two lattices are closer each other than in the individual lattices, this effect should play a central role in the study of twisted bilayers, which have been deeply studied in condensed matter physics by coupling hamiltonians of the equivalent continuous layers. In the domain of phononics, where the lattices are artificial man-made structures and contain much less scattering units, it is obvious that the individual analysis presented here will be the dominant phenomenon. Finally, we believe that these structures are promising candidates for the realization of mechanically tunable wave-trapping devices, not only for their efficient functionality but also for the relatively easy design based on multiple scattering theory. Although numerical examples have been reported for flexural waves in thin elastic plates, the conclusions of this work can be applied to any kind of waves, including electronic waves.

\begin{acknowledgments}
Daniel Torrent acknowledges financial support through the ``Ram\'on y Cajal'' fellowship under grant number RYC-2016-21188 and to the Ministry of Science, Innovation and Universities through Project No. RTI2018- 093921-A-C42. Marc Mart\'i-Sabat\'e acknowledges financial support through the FPU program under grant number FPU18/02725.
\end{acknowledgments}
%\bibliographystyle{apsrev}
%\bibliography{bibliography}

\end{document}